\documentstyle[11pt, aaspp4]{article}

\begin{document}

\title{Diagnostics of a Subsurface Radial Outflow From a Sunspot}

\author{D. C. Braun\altaffilmark{1}}
\affil{Solar Physics Research Corporation, 4720 Calle Desecada, Tucson, AZ,
85718}

\author{Y. Fan}
\affil{National Solar Observatory 950 N. Cherry Ave., Tucson, AZ 85719}

\author{C. Lindsey\altaffilmark{1}}
\affil{Solar Physics Research Corporation, 4720 Calle Desecada, Tucson, AZ,
85718}

\author{S. M. Jefferies\altaffilmark{1}} 
\affil{Bartol Research Institute, University of Delaware, Newark, DE,
19716} 

\altaffiltext{1}{Visiting Astronomer, National Solar Observatory, NOAO.
NOAO is operated by AURA, Inc. under contract to the National Science
Foundation}

\begin{abstract}
We measure the mean frequencies of acoustic waves
propagating toward and away from 
a sunspot employing a spot-centered Fourier-Hankel 
decomposition of $p$-mode amplitudes as measured from a 
set of observations made at the South Pole in 1991. 
We demonstrate that there is a significant frequency shift
between the inward and outward traveling waves which is consistent
with the Doppler effect of a radial outflow from the sunspot.
For $p$-modes of temporal frequencies of 3 mHz it is observed
that the frequency shift decreases slightly with spatial frequency,
for modes with degree $\ell$ between 160 to 600. From the 
$\ell$ dependence of the frequency shift, we
infer that the mean radial outflow within the observed annular
region (which extends between 30 and 137 Mm from the spot)
increases nearly linearly with depth, reaching a magnitude
of about 200 m/s at a depth of 20 Mm. 
This outflow exhibits properties similar to flows 
recently reported by Lindsey, et al. (1996)
using spatially sensitive local helioseismic techniques.

\end{abstract}


\clearpage
\section{Introduction}

In the past several years, observational and theoretical tools have
been developed in the field  of local
helioseismology to study the interaction
of solar acoustic waves with localized structures within the solar
interior. These have included the use of ``ring diagrams'' as probes of
local convective motions (Hill 1988, Patron, et al. 1995),
time-distance helioseismic measurements of sound speed and Doppler
perturbations (Duvall et al. 1996, D'Silva et al 1996, Kosovichev 1996),
scattering and absorption of waves by sunspots (Braun, Duvall and LaBonte 1988,
Bogdan, et al. 1993, Braun 1995, 
Fan, Braun and Chou 1995), and acoustic power maps (Lindsey and
Braun 1990, Braun, et al 1992, Toner and LaBonte 1993, Lindsey et al. 1996).

Recently, new evidence for subsurface flows associated with active regions
has been presented.
Duvall, et al. (1996) have used time-distance helioseismic
techniques to construct spatially resolved maps of the $p$-modes travel 
times over the solar surface, using full-disk solar images obtained at 
the geographic South Pole in 1991.  They found
considerable time travel decreases in acoustic rays propagating
away from active regions,
which they interpreting as being due to 2000 m/s downflows within
the upper 2 Mm of the convection zone.
Lindsey, et al. (1996) constructed acoustic power maps of the same
South Pole data, filtered in the Fourier domain to show Doppler 
signals caused by horizontal flows. These maps showed the presence
of outflows from active regions which, for mature regions
appeared to be predominately
concentrated at depths greater than 8 Mm below the surface.

An important analysis procedure employed in the exploration of
$p$-mode -- sunspot interactions has been the decomposition of
the solar oscillation signal, observed in an annulus surrounding
a spot, into appropriate inward and outward propagating wave modes.
This method has been termed ``Fourier-Hankel spectral decomposition'',
since the annulus is usually chosen to be small enough so that
the spatial (radial) form of the wave modes is described very closely
with Hankel functions. The initial utility of this procedure was demonstrated
by the exploration of the $p$-mode absorption qualities of sunspots and other
solar activity by looking at the difference in amplitudes between 
the inward and outward wave components (Braun, Duvall, and LaBonte 1988,
Bogdan et al. 1993, Braun 1995, Chen, et al. 1996). Braun, et al. (1992) 
and Braun (1995) have subsequently determined the scattering phase shifts
and mode-mixing amplitudes due to the refractive properties of spots.

We suggest that Fourier--Hankel spectral 
decomposition methods are highly useful in studying horizontal 
flows associated with active regions. For example, a net radial 
inflow or outflow centered on a particular point (e.g. a sunspot) will
produce equal but opposite shifts in the horizontal wavenumber (at constant
temporal frequency) of $p$-modes traveling toward and away from the center.
The signature of a radial
flow would then be a shift in the position of the $p$-mode ridge observed
in power spectra constructed alternately from the inward and outward
propagating waves. This procedure has a well known analogy in the
determination of internal solar rotation by the measurement of
frequency differences between prograde and retrograde
traveling global $p$-modes. Before proceeding, we point out
one important distinction between the use of the Hankel decomposition 
method in determining horizontal flows and its use in probing the
absorbing and refracting properties of solar active regions. 
The amplitude and phase differences which were the primary
diagnostics in the latter
analyses were used to probe absorption and scattering effects 
which were largely confined to a area {\it within the inner radius of
the annulus}.
In contrast, the radial Doppler diagnostic we
are proposing here is, to first order, only sensitive to the mean
radial flow of the medium {\it between the inner and outer radii 
of the annulus}. A consequence of this is that the magnitude of
the observed frequency shift will depend not only on the magnitude
of the true velocity flow, but also on the choice of the annulus
dimensions.

\section{Data Reduction}

The data used in this analysis consists of a subset of a larger
dataset obtained at the Geographic South Pole by the NSO-NASA-Bartol
group (Duvall, et al. 1996) between November 1990 and January 1991.
Ca II K-line images of the Sun during this time may be found in both the works 
cited above, as well as on the cover of the 1995 October issue of 
{\it Physics Today} (Harvey 1995).
For this study, we selected a mature active region, NOAA 6431,
consisting of a single, nearly circular sunspot (with umbral
and penumbral radii of 4 and 9 Mm respectively) surrounded by
a small region of plage. The data we utilize
consists of 185 hours of observations from Jan. 1-9, 
1991, with a 67\% data coverage.
As a control measure,
a region of quiet Sun centered at the same solar latitude
as the sunspot and offset 40 degrees east in longitude was
also selected for an identical analysis.

The data reduction procedure is 
described in detail in Braun (1995). Some additional considerations
regarding Fourier-Hankel spectral decomposition technique are given
by 
Braun, Duvall, and LaBonte (1988) and Bogdan, et al. (1993).
The K-line intensity values are first interpolated onto
a spherical polar coordinate system $(\theta,\phi)$
with the sunspot situated at $(\theta  = 0)$. The annular
region is defined by the inner and outer polar angles $\theta_{min}$
and $\theta_{max}$ respectively.
For values of $\theta \ll \pi $, we may employ Hankel functions as
approximations to the Legendre function decomposition. Thus, we decompose
the incident and scattered waves into components of the form
\begin{equation}
\label{eq1}\Psi _m(\theta ,\phi,t)=e^{i(m\phi +2\pi \nu t)}[A_m(\ell,\nu
)H_m^{(1)}(\ell\theta )+B_m(\ell,\nu )H_m^{(2)}(\ell\theta )], 
\end{equation}
where $m$ is the azimuthal order, $H_m^{(1)}$ and $H_m^{(2)}$ are Hankel
functions of the first and second kind respectively, $t$ is time, $\nu $ is
the temporal frequency and $\ell$ is the spatial wavenumber (which may
be compared with the degree of a spherical harmonic).
$A_m$ and $B_m$ represent the
complex amplitudes of incoming and outgoing waves respectively. 

The values of $\theta_{min}$ and $\theta_{max}$ should, strictly speaking, 
be determined from the actual radial size of the velocity features one
wishes to measure. Extending $\theta_{max}$, for example,
beyond the radial extent of flow adds no additional signal with the
disagreeable cost of increasing the level of noise. On the other hand, 
practical considerations suggest the annulus should not be too small
such that the spatial resolution is insufficient to isolate $p$-mode
ridges of adjacent radial orders. 
For the majority of
the measurements we present here we set $\theta_{min}$ and $\theta_{max}$
to 2.5 and 11.25 degrees (30 and 137 Mm) respectively, 
giving a resolution in $\ell$ 
of approximately 40. The outer radius of this annulus is comparable to
the horizontal extent of the outflows detected by Lindsey, et al. 
(1996).

The numerical transforms needed to compute the set of wave amplitudes 
$A_m(\ell, \nu)$ and $B_m(\ell, \nu)$ are described in Braun, Duvall, and 
LaBonte (1988). For each value of $\ell$, we measure the mode amplitudes 
for individual azimuthal orders for $|m|$ ranging from 0 up to a value
not to exceed $\ell\theta_{min}$, a condition required for the orthogonality
of the Hankel functions (Braun 1995). For the highest wavenumbers we 
employ azimuthal orders up to $m=20$.

\section{Results}

It is immediately apparent with only a visual examination of the 
power spectra that the ridges of the outgoing $p$-mode power are
shifted to higher temporal frequencies relative to the incoming 
waves. The frequency shift is more readily
visible when one corrects for the effects of $p$-mode absorption
by the sunspot by individually normalizing the incoming
and outgoing power.  Figure 1 shows the resultant power spectra of
two representative modes in the sunspot analysis.  
The large widths of the peaks are the result of a 
relatively poor wavenumber resolution.
In spite of this, frequency shifts
on the order of 10 $\mu$Hz are clearly seen between the incoming
and outgoing profiles. To measure the shift, we determine the 
power-weighted mean frequency for both power spectra. We define
$\Delta \nu$ as the mean frequency of the outward propagating mode
minus the mean frequency of the inward propagating mode.
Values of $\Delta \nu$ determined from both the sunspot analysis
and the quiet-Sun analysis are shown in the top and bottom panels
of Figure 2 respectively. For each value of $\ell$ we show the
frequency shift of the mode with temporal frequency closest to
3 mHz, which represent the most reliable determinations. Measurements
of $\Delta \nu$ of modes with frequencies below 3 mHz or above 4 mHz showed
considerable scatter in both the sunspot and quiet-Sun data.
Figure 2 clearly shows positive frequency shifts for almost all
modes analyzed in the sunspot data, while showing no detectable shifts
in the quiet-Sun analysis.  The most plausible cause of the frequency
shifts is a radial outflow from the sunspot. 

For values of $\ell$ below 240, the $p$-mode
ridges were too close together in the power spectra 
to measure the frequency shift using the annulus specified above.
The analysis was repeated using an annulus of twice
the width as the original annulus. This enabled the detection and
measurement of frequency shifts down to $\ell = 160$. A comparison of
the frequency shifts measured in both the smaller and larger annuli
for common modes showed that the shifts seen over the larger annulus
were on average a factor of 2.8 times smaller than observed with the
smaller annulus, suggesting that no flows are actually detectable
beyond 137 Mm, the outer radius of the smaller
annulus. The values of the shifts
for $\ell$ between 160 and 230 determined from the larger annulus and
multiplied by the factor 2.8 are shown as open circles in the
top panel of Figure 2.

Doppler frequency shifts of high degree p-modes produced
by horizontal flows have been previously studied 
(e.g. Gough and Toomre 1983, Hill 1988, Patron et al. 1995).
In cylindrical coordinates 
the presence of a radial axisymmetric flow $U_r \hat{\bf r}$ will
produce a relative frequency shift of
$ 2k \bar{U_r}$ between the incoming and outgoing
cylindrical waves, where $\bar{U_r}$ corresponds to an average
of $U_r$ over the annular region and depth weighted by the wave energy
density: 
\begin{equation}
\Delta \omega \equiv \omega^{out} - \omega^{in} = 2 k \bar{U_r}
= \frac{ 2 k \int_{0}^{\infty} ( \int_{r_1}^{r_2} U_r \, dr ) \;
K  \, d z }{ ( r_2 - r_1 )
\int_{0}^{\infty} K \, dz }
\end{equation}
where, $r_1$ and $ r_2$ denote respectively the inner and outer radii of
the annular region, $k$ is the horizontal wavenumber, $z$ denotes depth from 
the surface, and the kernel $K$ is the energy density of the particular wave
mode as a function of depth. In terms of our measured
frequency shift $\Delta \nu$ equation (2) can be expressed as
\begin{equation}
\Delta \nu \equiv \nu^{out} - \nu^{in}  
= \frac{ \ell \int_{0}^{\infty}  < U_r >  \;
K  \, d z }{  \pi R_{\odot} \int_{0}^{\infty} K \, dz }
\end{equation}
where $R_{\odot}$ is the solar radius and $< U_r >$ is the mean radial
flow over the annulus.

It is straightforward to see from equation (3) that a mean flow 
$< U_r >$ which is, say, constant with depth will produce frequency
shifts which are simply proportional to $\ell$.
The observed
behavior of the frequency shifts at 3 mHz (Figure 2), which
shows a slight decrease of $\Delta \nu$ with $\ell$,
implies therefore that the flows must actually {\it increase} with
depth. Using models of $< U_r >$ with different power law increases
with depth
$z$ we have computed numerically the expected frequency shifts
given by equation (3) for modes of $\nu$ = 3 mHz as a function of
$\ell$. The energy density kernels are computed from a standard solar
model (Christensen-Dalsgaard, Proffitt, and Thompson 1993).
We find that a mean velocity which increases as $z^{1.2}$  can 
explain the observed $\ell$ variation. The mean radial velocity is plotted
in the upper panel of Figure 3, and the calculated frequency shifts
are shown by the curves overlaying the data in Figure 2. The dashed
line indicates a mean radial velocity which continues to increase with 
a power law of exponent 1.2, while the solid line represents velocity which 
levels off to a constant value at a depth of 20 Mm below the 
surface. The modes with the lowest $\ell$ values observed penetrate 30 Mm
below the surface, so we are unable to infer 
velocities below this depth.

Having determined the radial outflow $<U_r>$ in the annular region
as a function of depth, we can estimate the vertical flow
using the requirement of the continuity
equation $\nabla \cdot ( \rho {\bf U}  ) = 0$, where $\rho$ is the density.
Consider the closed surface formed by an imaginary cylindrical tube
extending between the surface and a depth $z$ 
with its vertical axis centered on the sunspot and having a radius of
$r_1$ equal to the inner radius of the annular region.  
The mass flux through the top cross section at the surface is
naturally zero.  The mass flux flowing into the tube through the
bottom cross section is $\pi r_1^2 \rho (z) \, U_{vert} (z) $,
where $U_{vert}$ denotes the upflow velocity.
The mass flux flowing out of the cylinder tube through the tube
surface at $r_1$ is given by $2 \pi r_1
\int_0^z \rho (z') \, U_r (r_1, z') dz'$, where
$ U_r (r_1 , z)$ denotes the radial
outflow velocity at $r_1$. Here we approximate $U_r (r_1, z)$ by the
averaged $< U_r >$ in the annular region which is most likely a
lower estimate judging from the radial profile of the outflow obtained
by Lindsey et al.~(1996).  
Equating the flux through the bottom cross section to the flux through the
tube surface at $r_1$ as required by the continuity equation, we obtain
\begin{equation}
U_{vert} (z) = \frac{2}{r_1 \rho (z) } \int_0^{z} \rho \, < U_r > dz'.
\end{equation}
The lower panel of 
Figure 3 shows the resulting $U_{vert} (z)$, which is effectively an
averaged upflow velocity through the region defined by the inner annulus. 

Is it possible that the frequency shifts we are observing are due
to some effect which is not Doppler in nature?
We note that sunspots are efficient absorbers of acoustic
radiation and that a variation of the absorption with 
frequency will give rise to
a shift in the mean frequency of the outgoing mode peak. 
We have measured the absorption produced by NOAA 6431, which we find
to be similar to that exhibited by two sunspots analyzed in the
1988 South Pole data (see Figures 4 and 5 in Braun 1995). Of the
modes represented in Figure 2, those with $\ell$ below 350
show an increase of absorption with frequency along the ridge, the 
strongest variation amounting to a 5\% increase in the absorption coefficient 
over a 0.1 mHz frequency interval.  We have
estimated that this variation would produce a frequency shift, $\Delta \nu$,
at $\ell = 330$ of about $-2$ $\mu$Hz, which is of the 
opposite sign than that observed
and has a magnitude smaller than the formal errors shown in Figure 2.
The modes with $\ell >$  350 show essentially no variation in absorption 
along their respective ridges and are thus not affected 
in any detectable manner.

The outflow detected in NOAA 6431 by Hankel spectral decomposition techniques
appears very similar to the outflows 
detected by Lindsey, et al
(1996) using horizontal Doppler-sensitive acoustic power maps. 
The mean value of the radial velocity profile shown in Figure 5 of
Lindsey et al. over our annulus size is approximately 90 m/s, which 
is consistent with the flow speeds inferred here. Lindsey et al. also
find that for mature active regions the flows are predominately 
subsurface with speeds increasing with depth.

Mature sunspots are known to show surface outflows (``moats'')
detected by Doppler shifts and the proper motion of nearby magnetic features
(a recent summary is given by Brickhouse and LaBonte 1988). 
These flows have typical peak velocities
on the order of 500-1000 m/s and extend approximately 20 Mm beyond
the penumbra. 
The outflows detected using local helioseismology have 
significantly lower velocities,
and appear to persist to several times the radial extent of the moat.
It is possible that these outflows may represent a subsurface extension
of the moat flow. 

On the other hand, the upflow required to feed the subsurface outflow
appears to be at odds with the large downflows inferred by the time-distance
measurements of Duvall, et al (1996).  It is possible to produce wave 
travel time decreases along rays emanating from active regions with 
outflows and preliminary results using
time-distance methods which provide directional discrimination appear to support
this picture (Duvall, private communication).
However, velocities on the order
of 100 m/s appear to be an order of magnitude too small to match
the observed travel times.  We should not rule out the possibility of
rather complex flow patterns beneath active regions. A
shallow (2 Mm deep) eddy which is flowing down
at the border of the magnetic flux combined with a deeper
region of upflow which drives a subsurface outflow may 
reconcile the helioseismic data, although an additional
eddy at the extreme surface would be needed to
produce the moat flow. 

It is clear that the continued development and application of these 
and other observational 
techniques to the high quality data now becoming
available with the GONG and SOHO projects, combined with more sophisticated
theoretical modeling, will be crucial in 
exploring and understanding the subsurface structure 
and evolution of solar magnetic regions.

\acknowledgments

This research was supported by 
NASA grant NAGW--4143, NSF grants AST--9496171, ATM--9214714, 
OPP--9219515, DPP--8917626 and ONR Grant N00014-91-J-1040.

\clearpage

\begin{figure}
\plotone{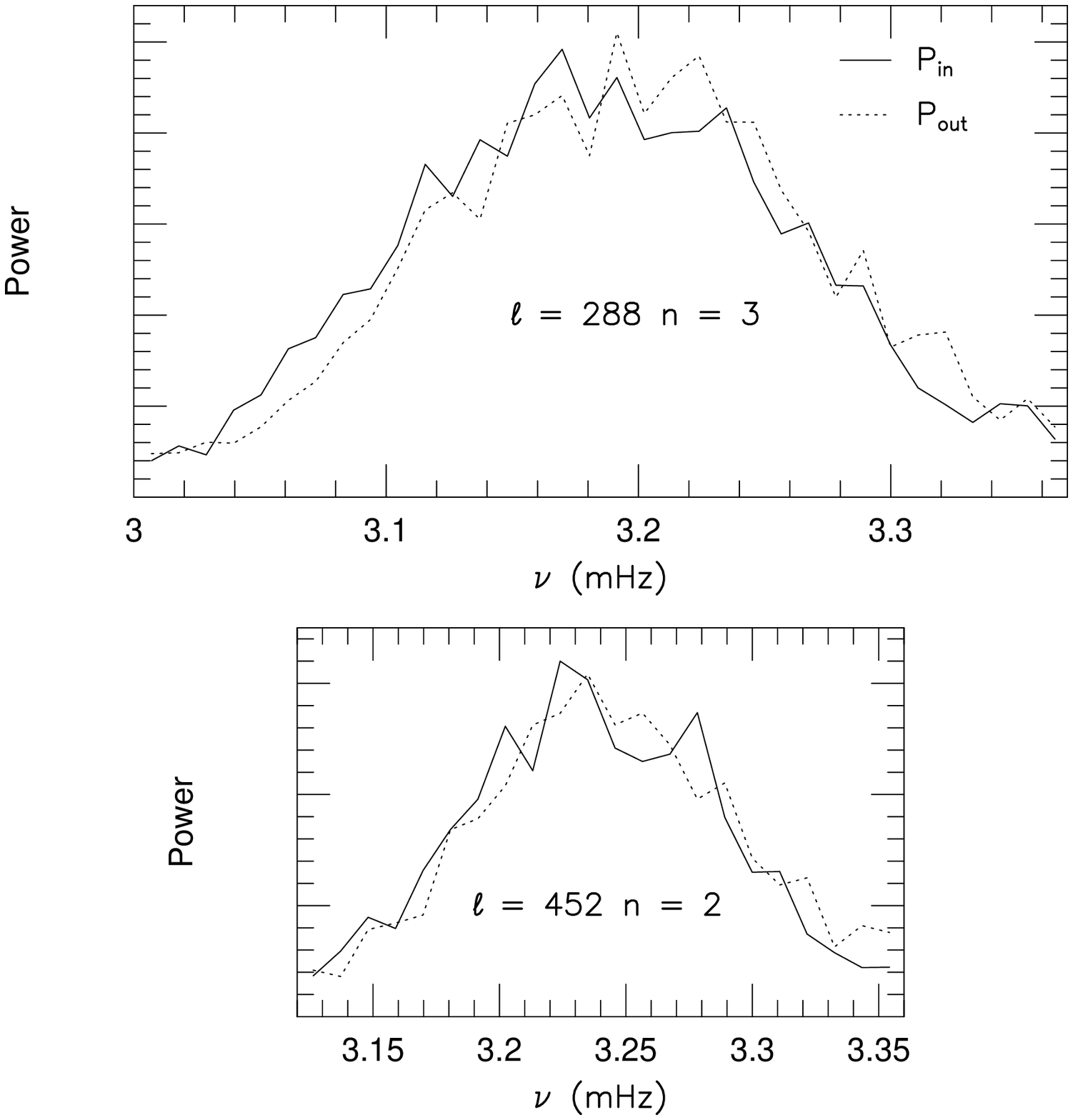}
\caption{
Power spectra corresponding to inward and outward propagating
$p$-modes for two representative modes. A relative shift in the peak of the 
outward modes towards higher frequency is apparent in both cases.
}
\end{figure}

\begin{figure}
\epsscale{0.8}
\plotone{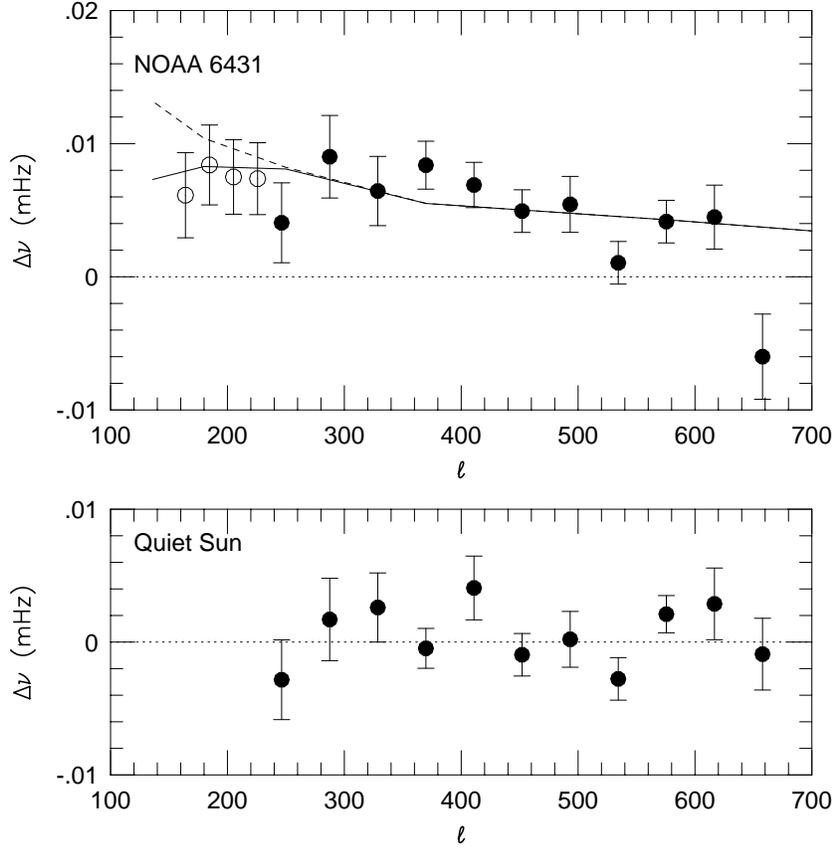}
\caption{
Measurements of the frequency shift plotted as a function
of degree $\ell$ for the sunspot (top panel) and quiet-Sun 
(bottom panel) analysis.
The filled circles (with formal 1 $\sigma$ error bars) represent 
measurements made over the small annulus (2.5 to 11.25 degrees from
the center) while the open circles in the upper panel represent
measurements made with a larger annulus (2.5 - 20 degrees), scaled
by a factor of 2.8. The curves show calculated frequency
shifts for two models of the mean radial outflow.
}
\end{figure}

\begin{figure}
\plotone{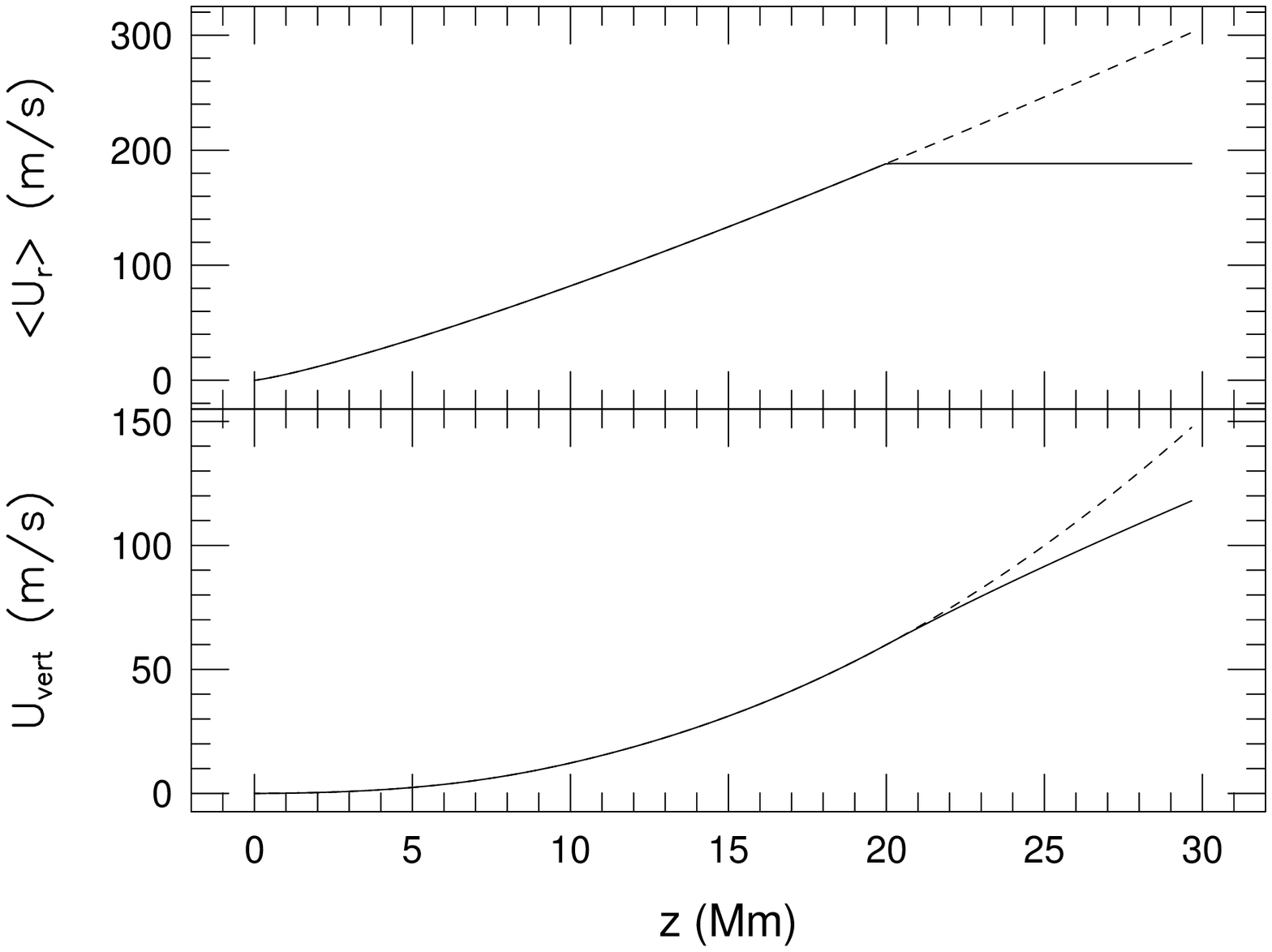}
\caption{
The top panel shows the inferred mean radial outflows
plotted against depth. The solid and dashed curves show the
outflows used to calculate the frequency shifts shown in corresponding
line types in Figure 2.  The bottom panel shows the upward 
vertical velocity inferred by flux conservation.
}
\end{figure}


\clearpage

\begin{references}

\reference{} Bogdan, T.~J., Brown, T.~M., Lites, B.~W., and Thomas, J.~H.
1993, ApJ, 406, 723

\reference{} Braun, D.~C. 1995, ApJ, 451, 859

\reference{} Braun, D.~C., Duvall, T.~L.~Jr., and LaBonte, B.~J. 1988, 
ApJ, 335, 1015

\reference{} Braun, D.~C., Lindsey, C., Fan, Y. and Jefferies, S.~M. 1992, 
ApJ, 392, 739

\reference{} Brickhouse, N.~S. and LaBonte, B.~J. 1988, Solar Phys., 
115, 43

\reference{} Chen, K.-R., Chou, D.-Y., and the TON team, 1996, ApJ, in press

\reference{} Christensen-Dalsgaard, J., Proffitt, C.~R., and Thompson, 
M.~J., 1993, ApJ, 403, L75

\reference{} D'Silva, S., Duvall, T.~L.~Jr., Jefferies, S.~M.,
Harvey, J.~W. 1996, ApJ submitted

\reference{} Duvall, T.~L.~Jr., D'Silva, S., Jefferies, S.~M.,
Harvey, J.~W., and Schou, J. 1996, Nature, 379, 235

\reference{} Fan, Y., Braun, D.~C. and Chou, D. 1995, ApJ, 451, 877

\reference{} Gough, D. O. and Toomre, J. 1983, Solar Phys., 82, 401

\reference{} Harvey, J. 1995, Physics Today, 48 (No. 10), 32 

\reference{} Hill, F. 1988, ApJ,  333, 996

\reference{} Kosovichev, A.~G. 1996, ApJ, in press

\reference{} Lindsey, C. and Braun, D. 1990, Solar Phys., 126, 101

\reference{} Lindsey, C., Braun, D.~C., Jefferies, S.~M., Woodard, M.~J., 
Fan, Y., Gu, Y., and Redfield, S. 1996, ApJ, submitted 

\reference{} Patron, J., Hill, F., Rhodes, E.~J., Korzennik, S.~G. 
and Cacciani, A.  1995, ApJ, 455, 746

\reference{} Toner, C.~G., and LaBonte, B.~J. 1993, ApJ, 415, 847

\end{references}
\end{document}